\documentclass[preprint2]{aastex} 

\shorttitle{Kaaret et al.}
\shortauthors{Chandra Observations of PSR B0540-69}

\begin{document}

\title{Chandra Observations of the Young Pulsar PSR B0540-69} 

\author{
P.\ Kaaret\altaffilmark{1}, 
H.L.\ Marshall\altaffilmark{2},
T.L.\ Aldcroft\altaffilmark{1}, 
D.E.\ Graessle\altaffilmark{1}, 
M.\ Karovska\altaffilmark{1},
S.S.\ Murray\altaffilmark{1},
A.H.\ Rots\altaffilmark{1}, 
N.S.\ Schulz\altaffilmark{2},
F.D.\ Seward\altaffilmark{1}
}

\email{pkaaret@cfa.harvard.edu}

\altaffiltext{1}{Harvard-Smithsonian Center for Astrophysics,
60 Garden St., Cambridge, MA 02138, USA}

\altaffiltext{2}{Center for Space Research, Massachusetts
Institute of Technology, Cambridge, MA 02139, USA}

\begin{abstract}

The young pulsar PSR B0540-69 was one of the first targets
observed with the Chandra X-Ray Observatory.  The high
angular resolution of Chandra allows us to resolve the
compact nebula surrounding the pulsar.  We have determined a
position for PSR B0540-69 of $\rm R.A. = 05^h 40^m 11^s.221,
decl. = -69^{\circ} 19\arcmin 54\arcsec.98$ (J2000) with a
$1\sigma$ radial uncertainty of $0.7\arcsec$.  Combining our
measurements of the pulsar period with previous measurements
covering a span of 12 years, we derive a braking index of
2.08.  The spectrum of the pulsed emission is consistent with
a power-law with a photon index of $1.83 \pm 0.13$.  The
compact nebula has a softer spectrum with a photon index of
1.85--2.26.

\end{abstract}


\section{Introduction}

PSR B0540--69 is a young, energetic pulsar located in the
Large Magellanic Cloud (LMC) and discovered in soft x-rays
using the {\it Einstein} observatory (Seward, Harnden, \&
Helfand 1984).  The pulsar is similar to the Crab pulsar in
period, 50~ms versus 33~ms for the Crab, characteristic
spin-down age, 1600~yr versus 1240~yr, and spin-down power,
$1.5 \times 10^{38} \rm \, erg \, s^{-1}$ versus $4.7 \times
10^{38} \rm \, erg \, s^{-1}$.  Comparison of detailed
observations of PSR B0540--69 versus those of the Crab should
be a useful test of models of young pulsars, their evolution,
and their associated nebulae.

Due to the large distance to the LMC, PSR B0540--69 is a
faint source.  While it has been been observed in the x-ray,
optical, and radio, relatively long observations on large
telescopes are required to obtain good signal to noise and,
thus, observational coverage is spotty.  Previous
observations of PSR B0540--69 have led to discordant
ephemerides and significant disagreements about the braking
index.

Here, we present an analysis of early Chandra X-Ray
Observatory observations of PSR B0540--69 made using both the
High Resolution Camera (HRC; Murray et al.\ 1997) and the
ACIS imaging camera (ACIS-I; Bautz et al.\ 1998).  An early
analysis of the HRC data has been previously described in
Gotthelf \& Wang (2000) which reported on the morphology of
the spatially resolved nebula near the pulsar and outer shell
of the remnant.  Here, we analyze the HRC data with an
improved aspect solution needed to obtain the correct
position for PSR B0540--69 and to remove artifacts present in
the initial processing of the data.  In addition, we present
results on the x-ray spectrum of the pulsar, the compact
nebula, and the outer remnant from an analysis of the ACIS
data.  We describe the observation and our analysis in \S 2. 
We present results on the source location, the pulsar period
history, the extent of the nebula, and the spectrum of the
emission in \S 3.  We conclude in \S 4.

\section{Observations and Analysis}

PSR B0540--69 was observed with the Chandra X-Ray Observatory
(CXO; Weisskopf 1988) during the verification and checkout
phase on 31 August 1999 using the High Resolution Camera
(HRC; Murray et al.\ 1997) for a total of 18~ks of good
observing time and on 26 August 1999 using the ACIS imaging
camera (ACIS-I) for a total of 20~ks of good observing time. 
All observations employed the High-Resolution Mirror Assembly
(HRMA; van Speybroeck et al.\ 1997).

The absolute time calibration of Chandra has not been
finalized.  For this reason, we use only frequency
measurements in the analysis below.  The accuracy of the
relative time-tagging within an observation is determined by
the spacecraft clock which is stable to better than one part
in $10^{9}$ over one day (Chandra Observatory Proposer's
Guide).  The conversion to a barycentric time frame
introduces an additional frequency uncertainty of less than
$2 \times 10^{-9}$ in these observations.  Thus, the accuracy
of the barycentric frequency determination is more than
sufficient for the frequency analysis presented below.

\subsection{HRC Analysis}

The HRC is a microchannel plate imager having very good
spatial and time resolution, but essentially no energy
resolution.  Each photon detected by the HRC is time tagged
with a precision of $16 \, \mu$s and position tagged with a
precision of 0.132\arcsec.  This makes possible time-resolved
imaging studies.  After launch, an error was found in the HRC
wiring which causes the event time tag to actually contain
the time of the previous event trigger (Seward 2000). 
Without correction, this leads to an timing error equal to
the time between successive event triggers.  The event
trigger rate is typically 250--300~s$^{-1}$, leading to a
typical error of 3--4~ms.  A special operating mode for the
HRC has been developed in which all event triggers are
telemetered to the ground and the timing error can be
eliminated by shifting of event time tags.  However, in the HRC
observation reported here, not all event triggers result in
valid events telemetered to the ground and correction of the
event times is not possible in all cases.  Thus, the event
times contain typical errors of 3--4~ms.  This error is small
relative to the 50~ms period of PSR B0540--69 and does not
significantly affect the period determination or the
pulsed-phased resolved imaging presented below.

This observation of PSR B0540--69 was only the third
observation made with the HRC of a celestial source and was
performed before the correct focus position of the HRC was
determined.  During this observation, the HRC was displaced
by $260 \, \mu \rm m$ from the true focus.  This defocuses
the image of an ideal point source, ignoring mirror and
aspect imperfections, to an annulus with inner and outer
radii of $0.17\arcsec$ and $0.33\arcsec$.

We applied aspect to X-ray events from the HRC using a
version of the Chandra X-Ray Observatory Center (CXC) aspect
pipeline (R4CU5, 11-Jan-2000) which utilizes elliptical
Gaussian centroiding.  Use of the improved aspect solution is
critical to obtain correct positions.  The relative aspect
reconstruction within the observation was checked by using
the aspect solution for this observation to de-dither the
optical guide star centroids.  This is similar to the process
used to apply aspect correction to x-ray events.  We found no
systematic residuals in the de-dithered star images, leading
us to conclude that relative aspect during the observation is
stable to within $0.3\arcsec$.  The absolute accuracy of the
aspect reconstruction was tested using a series of
observations of 8 HRC-I observations of known x-ray point
sources with identified optical counterparts with very
accurate astrometry (Aldcroft et al.\ 2000).  In each case,
the derived X-ray source position was compared with the known
counterpart position to derive the celestial location error. 
The absolute source coordinate uncertainty is $0.7\arcsec$
($1\sigma$ radial error).

The data were filtered using event screening techniques
(Murray et al.\ 2000) to eliminate ``ghost'' events produced
by the HRC electronics (Dobrzycki 2000).  Screening to
eliminate ``ghost'' events is particularly important due to
the brightness of PSR B0540--69.  The screening removes image
artifacts which could have been incorrectly interpreted as
features of the x-ray emission.

We extracted an image which shows the pulsar as a bright
point source surrounded by nebular emission with a roughly
elliptical shape with an extent of a few arc-seconds.  We
used centroiding as implemented in the standard Chandra
software routine celldetect to determine the position of the
point source (CIAO V1.1 Software Tools Manual).  The
coordinates found are $\rm R.A. = 05^h 40^m 11^s.168, decl. =
-69^{\circ} 19\arcmin 55\arcsec.12$ (J2000).

These coordinates were used to perform an initial barycentric
correction employing the JPL DE-405 ephemeris.  We searched
for pulsations using the $Z^2$ statistic also known as the
Rayleigh test (e.g.\ Buccheri, Ozel, \& Sacco 1987).  As the
pulse profile of PSR B0540--69 is known to be roughly
sinusoidal (Seward et al.\ 1984), we used only the first
order $Z^2$ statistic.  Pulsations were easily detected from
PSR B0540--69.  

As the diffuse emission from the nebula surrounding PSR
B0540--69 may affect the position determination, we created a
difference image to isolate the pulsed source. We made a
``pulsar-on'' image using photons taken from an interval of
0.5 in phase with an offset chosen to maximize the number of
photons selected, and a ``pulsar-off'' image from the
remaining data.  The ``pulsar-off'' image was subtracted from
the ``pulsar-on'' image to obtain the difference image.   
All x-ray sources, in particular the constant nebular
emission, not pulsed at the folding frequency are removed by
this procedure.  The difference image shows only a single
point source.

We note that the ``pulsar-off'' image does contain some
pulsar emission because the pulse profile is roughly
sinusoidal (Seward et al.\ 1984) and is not zero during the
full 0.5 in phase of the ``pulsar-off'' image.  The HRC
timing error causes some additional pulsar photons to be
shifted into the ``pulsar-off'' image, but the effect is
relatively small since the timing error of 3--4~ms is small
compared to the phase bin size of 25~ms.  The presence of
some pulsar emission in the ``pulsar-off'' image reduces the
amplitude of the pulsar peak in the difference image. 
However, we found that the difference image still contains a
highly significant peak at the pulsar position and that use
of the maximum size phase bins was advantageous in reducing
image noise away from the peak where the photon statistics
are much lower.  We also note that the timing error produces
a very small net shift of nebular photons into the
``pulsar-on'' image.  If the timing error were simply a
random number added to the time tag, then the net shift would
be zero since the number of nebular photons moved from the
``pulsar-off'' image into the ``pulsar-on'' image would equal
the number moved in the opposite direction.  However, since
the error is actually a shift of time tags, the fact that the
event trigger rate is slightly higher during the
``pulsar-on'' interval (by a factor equal to the pulsar count
rate divided by average total event trigger rate or roughly
0.3\%) produces a net shift.  We estimate that the net shift
of nebular photons into the ``pulsar-on'' image is less than
0.05\%.

Using the position obtained from this difference image, we
performed a second barycentric correction and selected events
within a circle with a radius of 9 HRC pixels (1.1\arcsec)
around the difference image position.  We again searched for
pulsations and found a strongly significant pulsation signal
at 19.7988001(21)~Hz at epoch MDJ 51421.6240; the digits in
parentheses indicate the $1\sigma$ uncertainty.

\begin{figure}[tb] \epsscale{0.9} \plotone{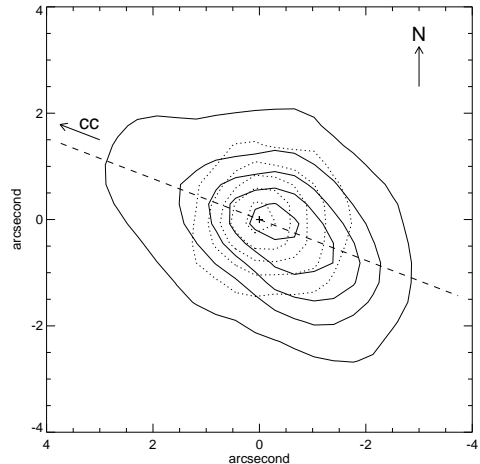}
\caption{Contour plots of emission from PSR B0540--69.  The
solid contours are the ``pulsar-off'' phase interval and the
dotted are the difference image.  The cross is the pulsar
position.  Both images were smoothed with a Gaussian with
FWHM = 0.7\arcsec.  The contour levels are 0.1, 0.3, 0.5,
0.7, and 0.9 of the peak pixel.  The dashed line and arrow
labeled ``cc'' indicate the orientation of the ACIS image.}
\label{0540image} \end{figure}

We made a second difference image using this best period, see
the dotted contours in Fig.~\ref{0540image}.  The position of
the source in the two difference images agree within $\rm 0.3
\, pixel = 0.04\arcsec$.  The position from the final image 
is $\rm R.A. = 05^h 40^m 11^s.221, decl. = -69^{\circ}
19\arcmin 54\arcsec.98$ (J2000).  The position uncertainty is
dominated by the accuracy of the aspect reconstruction which
we take to be $0.7\arcsec$ ($1\sigma$ radial error).  For the
difference image, 50\% of the flux is contained with a 
diameter of 1.3\arcsec.  This is significantly larger than
the half-power diameter of $0.76\arcsec$ measured for AR Lac
in calibration observations made to determine the on-orbit
point spread function (Jerius et al.\ 2000), but is
consistent with the expected combined effects of the HRMA
point-spread function given the harder spectrum of PSR
B0540--69, the relative aspect accuracy measured for this
observation, the position resolution of the HRC, and the
defocusing.

\subsection{ACIS Analysis}

The ACIS-I camera consists of an array of 4 front-illuminated
charged coupled devices (CCDs).   The physical pixel size is
0.24 $\mu$m which at the aim-point of the telescope is
comparable to the $0.5\arcsec$ spatial resolving power of the
HRMA.  Each CCD contains $1024 \times 1024$ pixels organized
into 4 readout nodes each of which reads out 1024 rows and
256 columns of pixels.  The pulsar was positioned at the
aimpoint of the telescope, which is about 960 rows away from
the readout node on ACIS-I device I3.  The pulsar and its
surrounding nebula fit very well onto node 3 of device I3,
therefore we use only this node in our analysis.  The ACIS
flight software was set to an on-board processing mode that
retains the $3 \times 3$ pixel event island.

In order to have sufficient time resolution to resolve the
pulses from PSR B0540--69, the ACIS-I was operated in
continuous clocking mode (``cc-mode'').  In this mode, charge
is continually shifted between successive pixels in each row
and the pixel at the base of each row is read out after each
charge transfer.  This leads to acquisition of a
one-dimensional image each 3.1~ms. The 1-d image integrates
flux along the readout direction.  The time of arrival of
each photon must be calculated from the time of readout and
the (unknown) position of arrival along the readout direction
divided by the speed of pixel transfer along the readout
direction.

The data on ground were manually processed through the
standard CXC pipelines into CXC level 1 event lists.  As
software for the analysis of cc-mode data is not currently
available from the CXC, we further processed the ACIS-I data
using custom IDL procedures.  The pointing direction was
transformed to the spacecraft coordinate frame ($s_y, s_z$,
in detector pixels) so that the event positions could be
corrected for dither in the detector $x$ direction only via
$x_p = x_t + S*\delta x(t_i) + x_r$, where $x_p$ is the
position of the event projected along a line on the sky (at a
position angle of 66.97\arcdeg), $x_t$ is the event
coordinate in ``tiled'' coordinates ({\bf tdetx}), $S$ is the
sign of the correction for dither (-1 for chips I1 and I3 and
+1 for chips I0 and I2), and $x_r$ is a random variable
between -0.5 and 0.5 that reduces effects of aliasing.  The
spacecraft dither and optical bench distortions are accounted
in the term $\delta x(t)$, which is the derived from
interpolating $\delta x(t) = s_z(t) - dz_{SIM}(t)$, where
$dz_{SIM}(t)$ gives the motion of the science instrument
module (SIM) measured by the on-board fiducial lights.  

\begin{figure}[tb] \epsscale{0.7} \plotone{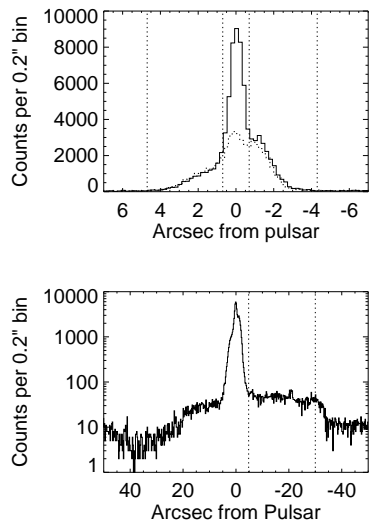}
\caption{One-dimensional images from ACIS-I cc-mode data. The
top panel shows two image profiles.  The solid curve shows
events with pulse phases between 0.1 and 0.3 where the pulse
is maximum and the dashed curve shows phases 0.6 to 0.8 where
the pulse is at minimum.  The bottom panel shows the overall
profile.  The dotted lines indicate the various regions used
in the spectral analysis.} \label{ccimage} \end{figure}

The histogram of the $x_p$ values, i.e. the one-dimensional
image perpendicular to the readout direction, is shown in
Fig.~\ref{ccimage}.  For this observation, the 1-d image is
along a positive angle of 67\arcdeg, or approximately ENE, as
indicated on the HRC image in Fig.~\ref{0540image}.  We
define the angle along the 1-d image as positive to the ENE. 
A reference point is computed by fitting a narrow Gaussian to
the brightest pixels, assuming that the synchrotron nebula
provides a simply sloping background.  The profile of the
pulsar is well fitted by a Gaussian with a dispersion of
0.30\arcsec, which is consistent with the telescope point
spread function (PSF).  The coordinates in
Fig.~\ref{ccimage} are relative to the fitted peak
position.  The sharp peak of the pulsar is flanked by the
integrated image of the compact nebula.  The compact nebula
brightness distribution peaks to the SW from the pulsar
position and declines rapidly further SW after the peak.  The
brightness falls off more slowly toward the NE.  The compact
nebula gives way to a more diffuse remnant, the outer
supernova remnant, at about $5\arcsec$ from the pulsar.  The
diffuse emission extends to $\sim 35\arcsec$ in one direction
and is truncated by a chip gap at $50\arcsec$ from the pulsar
in the other direction.

The event times were constructed using the event row numbers,
$y_i$, an interpolation of the dither and SIM correction,
$\delta y(t) = s_y(t) - dy_{SIM}(t)$; the ACIS exposure time
to UT lookup table, $t_{UT}(e)$; the geocenter and solar
system barycenter corrections, $dt_{\earth}$ and $dt_{\sun}$;
and a random variable, $y_r$, between 0 and 1 that reduces
effects of aliasing: $t_{TT,i} = t_{UT}(e_i) + t_f(y_i + y_r)
+  (t_{TT} - t_{UT})$, where the last term is simply a
constant for this observation (64.184~s), $t_f$ is the
average detector frame shift time in ephemeris time units
(about 0.00284996~s).  The exposure time to UT lookup table
should be accurate to less than 1 msec.  Finally, the times
corrected to the solar system barycenter (TCB) are $t_i =
t_{TT,i} - \delta y(t_{TT,i}) + dt_{\earth}(t_{TT,i}) + 
dt_{\sun}(t_{TT,i})$.  

We have found that an additional 5~s must be added to
$t_{UT,i}$ in order to match the TT times computed for the
aspect data.  The 5~s offset is a data processing anomaly
that has not been explained fully so we merely report that it
exists.  Specifically, the offset is needed to bring the
time-tagged aspect data into concurrence with the time-tagged
photon events.  Since we believe that the aspect data have
``correct'' times (based on data processing of other
instrument configurations), we have adjusted the event
times.  We can state that this 5~s offset is good to about
1~s and that it has not changed across the observation at
this level.  More importantly, different cc-mode
observations, even $\sim 50$~days later, give the same value
of the shift to within 1~s, thus the secular variation in the
offset is no larger than $3 \times 10^{-7}$.  If there were a
secular drift of this timing offset at this level, then it
would produce a 5~ms timing error across the ACIS
observations and lead to an error on the period roughly twice
the statistical error quoted below.  We use this larger error
in the ephemeris fit described below.  The barycenter timing
correction changes by only 60~ms across the ACIS observation
and is approximately quadratic in form, thus a 5~s timing
uncertainty produces an uncertainty of less than $1.4 \times
10^{-9}$ on the barycenter frequency correction.

The events within $0.7\arcsec$ of the peak in the $x_p$
distribution were Fourier transformed to obtain the candidate
pulse frequency and the best estimate was determined from
pulse folding: 19.7988900(30) Hz.  Two imaging profiles were
obtained by selecting from phases 0.6 to 0.8 where the pulse
is at a flat minimum, and phases 0.1 to 0.3 where the pulse
is nearly level at maximum.  These two image profiles, along
with the profile of all the emission, are shown in
Fig.~\ref{ccimage}.

\subsubsection{Spectral analysis}

The observation took place in a period where the ACIS
instrument experienced large changes in the spectral response
due to low energy cosmic proton impacts.  This resulted in a 
significantly increased charge transfer inefficiency (CTI). 
This has two effects on the spectral response.  First, the
energy scale changes due to charge loss that is not recovered
during the readout, and, second, the introduction of
additional electron noise due to enhanced charge trapping
worsens the spectral resolving power.

\begin{figure}[tb] \epsscale{0.9} \plotone{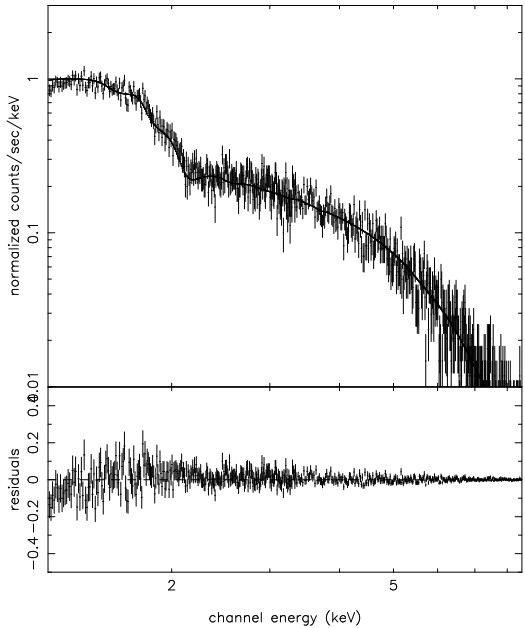}
\caption{Spectrum from PSR B0540--69.} \label{psr_spec}
\end{figure}

The focal plane temperature of the device  at the time of the
observation was -100$^{\circ}$ C.  Unfortunately, during this
period of change there are no response matrices available at
a focal plane temperature of $-100^{\circ} \rm \, C$. 
However, there are calibration products available for a focal
plane temperature of $-110^{\circ} \rm \, C$.  Calibration
data indicate that throughout most of the energy scale the
change in CTI with focal plane temperature is fairly linear. 
The change in spectral resolution is then assumed to be
linear as well. Calibration data also indicate how CTI
changed during the period  of degradation.  By estimating the
change in energy scale due to CTI as well as taking in
account the energy scale shift implied by the temperature
change from $-100^{\circ} \rm \, C$ to $-110^{\circ} \rm \,
C$, we find that the detector area between row 100 and 320 at
$-110^{\circ} \rm \, C$ matches the energy scale of the
detector node at the time of the observation between rows 800
and 1024 at $-100^{\circ} \rm \, C$ quite well.  In order to
fine tune the energy scale, we made fits using the available
response matrices at $-110^{\circ} \rm \, C$ valid for rows
100 to 320 and selected the ones that matched the expected 
mirror Ir-edge structures and position with the one expected
for the model effective area.  Currently available response
matrices show a quite strong non-linearity in the energy
scale below 1 keV that can affect the analysis of the soft
part of the spectrum.  For the analysis here, we do not fit
the spectra below 1.2~keV.  Fig.~\ref{psr_spec} shows a fit
of the pulsar spectrum (see below).  By examining known edges
in the CCD response, particularly the Si K edge at 1.738~keV,
we estimate that the energy scale is accurate to  within 2\%.

Background analysis is important in cc-mode because each
pixel read out has the integrated background from 1024 CCD
pixels.  The contribution does not exceed 2\% for the pulsar
or its compact nebula, but is almost 30\% for the outer
remnant.  We use the source free region between $-35\arcsec$
and $-50\arcsec$ to compute the background spectrum after
cleaning the event list for all apparent hot columns. The
measured average background rate was $1.20 \pm 0.01 \times
10^{-3} \rm \, cts \, s^{-1} \, column^{-1}$.

For the effective area we used the one available for the
standard ACIS grade set, which is appropriate since the
observation was performed in a mode that allowed
reconstruction of a $3 \times 3$ pixel island for each event.
In order to compute the corresponding ancillary data file
(arf) we used custom software that averaged over the aspect
solution.  This is legitimate for our analysis because no
source portion in the analysis crosses a CCD or node
boundary.  The spectral fits were performed with XSPEC
v.10.0.

\begin{figure}[tb] \epsscale{0.9} \plotone{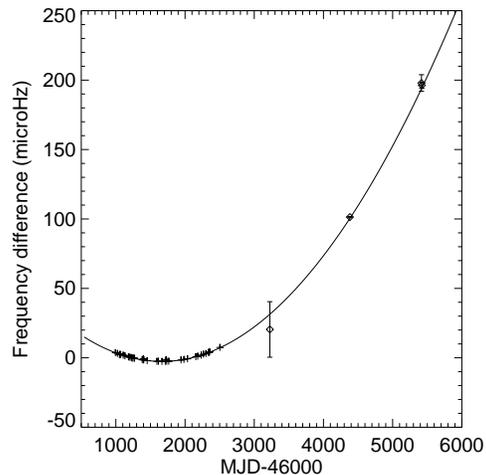}
\caption{Deviation of frequency from a linear ephemeris.  The
crosses indicate data from Deeter et al.\ (1999) and the
solid line is the corresponding quadratic ephemeris.  The
four diamonds indicate more recent data from, in
chronological order, Boyd et al.\ (1995), Mineo et al.\
(1999), and the ACIS and HRC values from this work.}
\label{fitnu} \end{figure}

\begin{deluxetable}{lr}
\tablecolumns{2}
\tablecaption{Spin Parameters for PSR 0540--69.\label{ephem}}
\tablewidth{0pt}
\tablehead{\colhead{Parameter} & \colhead{Value}}
\startdata
$t_{0}$ (MJD) &  47700.0 \\
$\nu_0$ (Hz) & 19.8593584982(40) \\
$\dot{\nu}_0$ ($10^{-10} \rm \, Hz \, s^{-1}$) & -1.8894081(7) \\
$\ddot{\nu}_0$ ($10^{-21} \rm \, Hz \, s^{-2}$) & 3.7425(43) \\
$n$ (braking index) & 2.0820(24) \\
\enddata
\vspace{-0.1in} \tablecomments{Numbers in parentheses are $1 \sigma$ errors
in the last quoted digits.}
\end{deluxetable}

\section{Results and Discussion}

\subsection{Timing}

In Fig.~\ref{fitnu}, we compare our two period measurements
and two other recent observations of PSR B0540--69 (Boyd et
al.\ 1995; Mineo et al.\ 1999) with the ephemeris derived by
Deeter et al.\ (1999) using data from MJD 46992-48335.
Extrapolation of the Deeter et al. (1999) ephemeris shows
that the new data, which extend almost 8 years beyond the
data used to derive the ephemeris, differ by at most
$2\sigma$ from the predicted frequencies.  We interpret this
as a strong indication that the timing solution found by
Deeter et al.\ (1999) is accurate.  Both Chandra frequencies
are consistent, within the $1 \sigma$ errors, with the
extrapolation of the Deeter et al.\ (1999) ephemeris.  The
HRC frequency measurement is within $1.9 \times 10 ^{-6} \rm
\, Hz$ of the predicted value and suggests that no large
persistent offsets in pulse frequency or its first derivative
occurred during the 8 year interval.  However, we are unable
to rule out small persistent offsets $\Delta \nu < 2 \times
10^{-6} \rm \, Hz$ or $\Delta \dot{\nu} < 8 \times 10^{-15}
\rm \, Hz \, s^{-1}$, similar to those reported by Deeter et
al.\ (1999).

We combined our frequency measurements with those of Boyd et
al.\ (1995), Mineo et al.\ (1999), and Deeter et al.\ (1999;
specifically the frequency data used to derive the ephemeris
and marked with note~5 in Table~4) to derive the ephemeris in
Table~\ref{ephem}. The ephemeris is consistent, within
errors, with that of Deeter et al.\ (1999).  We find a
braking index of $n = 2.0820 \pm 0.0024$ ($1\sigma$
uncertainty) at an epoch 47700~MJD.  This braking index is
much lower than the index of 3 expected for energy loss via
electromagnetic dipole radiation and is significantly lower
than the indices measured for the Crab pulsar or PSR
1509-58.  The low value of the braking index may be due to a
pulsar wind (Manchester \& Peterson 1989; Nagase et al.
1990). The synchrotron nebula may provide information on the
properties of the pulsar wind and it would be of interest if
the properties of nebulae surrounding young pulsars could be
related to their braking indices.  Other possible reasons for
the low value of the braking index are distortion of the
magnetic field lines (Manchester \& Taylor 1977), or a
time-varying magnetic field strength (Blandford \& Romani
1988).  Continued pulse timing of PSR B0540--69 should make
it possible to distinguish amongst these alternatives.

\subsection{Position}

Positions for PSR B0540--69, or its optical counterpart, have
been reported by Seward et al.\ (1984) from X-ray imaging, by
Deeter et al.\ (1999) from X-ray and optical timing, by
Caraveo et al.\ (1992) and Schmidtke et al.\ (1999) from
optical imaging, and by Shearer et al.\ (1994) from
time-resolved optical imaging.  Our position is in good
agreement with those of Schmidtke et al.\ (1999) and Shearer
et al.\ (1994) and marginally consistent with that of 
Caraveo et al.\ (1992).  The results of Shearer et al.\
(1994) indicate that the pulsar may be slightly displaced
from the maximum in the steady optical emission identified
with the pulsar by Caraveo et al.\ (1992) in a direction
which would improve the agreement with our position
measurement.  Our HRC position for PSR B0540--69 lies
2.6\arcsec\ from the original X-ray position reported by
Seward et al.\ (1984), although within the 90\% confidence
error box.  Our position is $1.8\arcsec$ from the position
derived from X-ray timing measurements with the Ginga
satellite by Deeter et al.\ (1999).  Interestingly, the
Deeter et al.\ (1999) position agrees very well with the
other positions determinations in declination, but is
significantly displaced from all the other positions in R.A. 
The location of PSR B0540--69 less than $4 \arcdeg$ from the
ecliptic pole may lessen the accuracy of the determination of
its R.A.\ from pulse timing.

Because a shift of 2.6\arcsec\ in source position will induce
an annual sinusoidal variation in the barycentric pulse
arrival times with an amplitude of several milliseconds, an
accurate position is critical when performing a pulse time of
arrival (TOA) analysis.  Most timing studies of PSR B0540--69
have used the original X-ray position and should be
reconsidered in light of the more accurate position reported
here.  In particular, the previous position uncertainty may
have contributed to the disparate range of braking indices
found for PSR B0540--69.  Also, the decreased uncertainty in
the position may improve the evidence presented for the
detection of timing noise in PSR B0540--69 (Eikenberry,
Fazio, \& Ransom 1998).

\subsection{Nebula}

We found a $0.35\arcsec$ displacement between the centroid of
the total x-ray emission near PSR B0540--69 and the pulsar
itself.  This relative displacement is significant and
represents a physical offset between the pulsar and the peak
x-ray emission in the nebula similar to the offset seen in
the Crab nebula versus pulsar.

The difference image, which isolates the presumably
point-like pulsar emission, provides an indication of the
point-spread function for a point source including the finite
resolution of the HRMA and HRC, the effects of the aspect
errors, and the focusing error described above.  Comparison
of the difference image and the ``pulsar-off'' image, which
highlights the nebular emission by reducing -- but not fully
eliminating -- the pulsar emission, shows that the nebula is
spatially extended and has a centroid which is slightly WSW
from the pulsar, see Fig.~\ref{0540image} in which the dotted
contours represent the difference image and the solid
contours represent the ``pulsar-off'' image.  

\begin{figure}[tb] \epsscale{0.9} \plotone{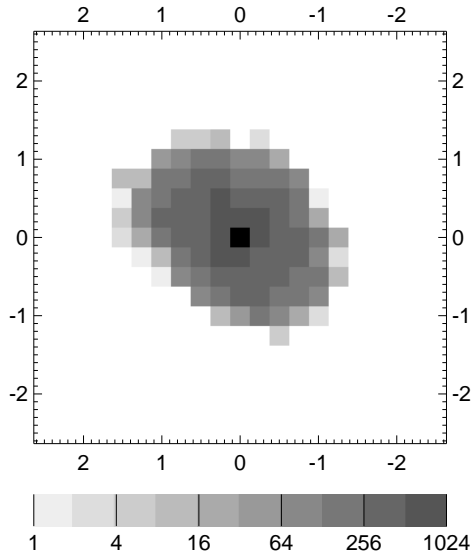}
\caption{Deconvolved image of the emission from PSR
B0540--69.  North is up and the axes indicate displacement
from the central pixel in arcseconds. The pixel size is
$0.26\arcsec$.  The scale indicates counts per pixel.  The
central pixel is saturated with 3230 counts.}
\label{0540decon} \end{figure}

To estimate the true extent of the nebula, we performed an
image deconvolution using the difference image described
above to represent the telescope point-spread function
including the effects of aspect errors and defocusing.  We
deconvolved an HRC image binned to 0.26\arcsec\ resolution
using the standard Richardson-Lucy technique (Richardson
1972; Lucy 1974).  Application of the deconvolution and also
the event screening described above is essential to determine
the true angular extent of the nebula.  The resultant
deconvolved image in shown in Fig.~\ref{0540decon}.  The
nebula has an angular diameter of $2\arcsec-3\arcsec$ and is
more extended along a NE-SW axis.

Gotthelf \& Wang found the same orientation for the compact
nebula, but a larger size.  Their results are consistent with
our un-deconvolved contour plot in Fig.~\ref{0540image}.  The
deconvolution gives a better estimate of the true angular
extent of the nebula.  Gotthelf \& Wang (2000) also report
marginal evidence for a ``jet-like feature'' extending to the
NW from the pulsar.  A similar weak feature is present in our
data.  However, the feature is present in the ``pulsar-on''
and difference images, but is  significantly weaker in the
``pulsar-off'' image.  Caution is required before
interpreting this as a physical feature and another
observation, with a different roll-angle, would be required
for confirmation.

The total HRC count rate from the nebula and pulsar within a
$4\arcsec$ radius is 0.84 counts s$^{-1}$.  Assuming a
power-law spectral model with a photon spectral index of 2.0
and an interstellar absorption column density $N_{H} = 4.6
\times 10^{21} \, \rm cm^{-2}$ (see the spectral fitting
below) gives an unabsorbed flux of $5.8 \times 10^{-11} \rm
\, erg \, cm^{-2} \, s^{-1}$ in the 0.2--10~keV band.

\subsection{Pulsed emission}

The HRC count rate within $1.1\arcsec$ of the pulsar is 0.38
counts s$^{-1}$ with 41\% of the signal pulsed.  For a
narrower radius of 0.6\arcsec, the pulsed fraction increases
to 55\%; however, some of the pulsed emission is lost due to
the large PSF during the HRC observation.  We note that the
measured pulse fraction is sensitive to the shape of the
pulse near the maximum and minimum, which may be affected by
the 3--4~ms timing error in the HRC.  Thus, the true pulsed
fraction may be somewhat higher than this value.

This pulsed fraction is significantly higher than previously
reported, e.g. $\sim 15\%$ from ROSAT (Finley et al.\ 1993),
because Chandra can resolve the surrounding compact nebula. 
Using a power-law spectral model with a photon spectral index
of 1.83 and a column density $N_{H} = 4.6 \times 10^{21} \,
\rm cm^{-2}$ (see the spectral fitting below) gives an
unabsorbed pulsed flux of $1.1 \times 10^{-11} \rm \, erg \,
cm^{-2} \, s^{-1}$ in the 0.2--10~keV band.  The off-phase
deconvolution (not shown here) shows only a weak point source
at the pulsar position, indicating that, like the Crab
Pulsar, any steady emission from the pulsar itself is small
compared with the pulsed component.

\begin{figure}[tb] \epsscale{0.9} \plotone{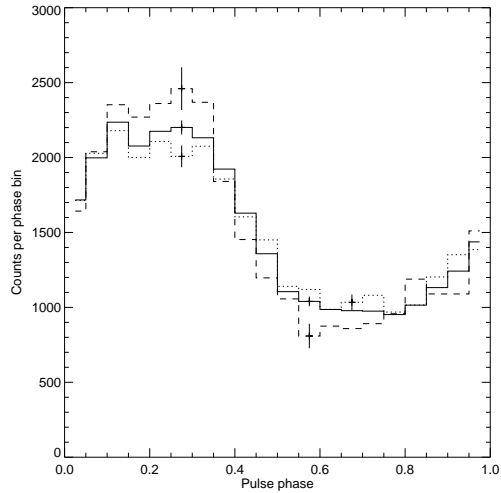}
\caption{Pulse phase histogram of the events from PSR
B0540--69 in the ACIS-I data.  The solid line indicates
photons within $0.7\arcsec$ of the pulsar position.  The
dashed line is for photons with energies above 4~keV and the
dotted line is for photons with energies below 1.5~keV; both
normalized to have the same total number of counts as in the
full band.  Representative error bars are shown.}
\label{cc_phasehist} \end{figure}

The ACIS pulse profile for photons within $0.7\arcsec$ of the
pulsar position is shown in Fig.~\ref{cc_phasehist}.  The
ACIS cc-mode data suffer from a significant nebular
contamination of the pulsar signal due to the integration of
counts along the readout direction.  Using the analysis of
the position profile described above, we estimate the nebular
contribution to be 570 counts per phase bin, leading to a
pulse fraction of $62 \pm 5\%$.  This estimate is consistent
with that found from the HRC data.  Fig.~\ref{cc_phasehist}
also shows the pulse profile for photons with energies above
4~keV and below 1.5 keV.  The modulation increases with
energy indicating that the pulsed emission is harder than the
unpulsed or nebular emission. The pulsed fraction, even below
1.5 keV, is significantly higher than reported from Rosat
(Finley et al.\ 1993) or BeppoSAX (Mineo et al.\ 1999) due to
nebular contamination in the previous results.

\begin{deluxetable}{lccc}
\tablecolumns{4}
\tablecaption{Parameters of Spectral Fits. \label{specfits}}
\tablewidth{0pt}
\tablehead{      & \colhead{Index} & \colhead{Flux} &
\colhead{$\chi^2_{\nu}$}}
\startdata
  Pulsar all     & 1.92$\pm$0.11   & 1.78$\pm$0.05  &   0.99 \\
  Pulsar ``off'' & 2.09$\pm$0.14   & 0.21$\pm$0.08  &   0.60 \\
  Pulsar ``on''  & 1.88$\pm$0.11   & 0.74$\pm$0.04  &   0.75 \\
  Pulsed         & 1.83$\pm$0.13   & 0.53$\pm$0.06  &   0.75 \\
  Nebula (left)  & 1.96$\pm$0.11   & 0.98$\pm$0.01  &   1.14 \\
  Nebula (right) & 2.12$\pm$0.14   & 0.49$\pm$0.01  &   0.63 \\
\enddata
\tablecomments{This Table contains the photon index from the
power-law spectral fit, the absorbed flux in the 0.6-10 keV
band in units of $10^{-11} \rm \, erg \, cm^{-2} \, s^{-1}$,
and the $\chi^2_{\nu}$ of the spectral fit.  The errors are 90\%
confidence.}
\end{deluxetable}

\subsection{Spectra}

We accumulated ACIS spectra from the pulsar, two regions
surrounding the pulsar where the compact nebula is prominent,
and the outer region where the outer remnant dominates. 
These spectra were analyzed using the response matrices
described above.  In this section, we quote x-ray fluxes for
the energy band 0.6--10~keV and errors for 90\% confidence
unless otherwise noted.

\subsubsection{Pulsar spectrum}

For the pulsar emission we selected events within 0.7\arcsec\
of the peak position in Fig.~\ref{ccimage}.  The background
subtracted source rate was 1.49 cts/s. Pile-up here is
negligible because of the small clocking time in cc-mode. We
used this overall spectrum to determine the proper  detector
response matrix for the analysis.  Since we ignore energies
below 1.2 keV, we are less able to constrain the column
density, $N_H$, in our fits. In order to fit all spectra
equally well, it was best to fix $N_H = 4.6 \times10^{21} \,
\rm cm^{-2}$.  The overall spectrum (see
Table~\ref{specfits}) is well represented by a power law of
index $1.92 \pm 0.11$.  Both $N_H$ and the photon index are
in good agreement with the values found by Finley et al.\
(1993) with the Rosat PSPC and by Mineo et al.\ (1999) with
BeppoSAX.

In order to measure the spectrum of the pulsed emission
alone, we have to subtract the background imposed by the
emission of the nebula.  This can be done by making use of
the pulse phase resolved components.  We accumulated spectra
at pulse phases 0.6--0.8 when the pulsed emission is at a
minimum, the ``pulsar-off'' spectrum, and at pulse phases
0.1--0.3 when the pulsed emission is maximum, the
``pulsar-on'' spectrum.  We subtracted the ``pulsar-off''
spectrum from the ``pulsar-on''spectrum to produce the pulsed
emission spectrum.  Table~\ref{specfits} shows the result of
the spectral fits.  The ``pulsar-off'' spectrum has a
somewhat softer spectrum than the ``pulsar-on'' spectrum. 
The pulsed emission spectrum shows an even harder spectral
index of $1.83 \pm 0.13$.  The total absorbed pulsed flux can
be calculated by integrating the emission above the lowest
pulse phase.  If we consider the flux level at phase 0.7 as
the unpulsed flux level, we find a pulsed flux of $7.4 \pm
0.2 \times 10^{-12} \rm \, erg \, cm^{-2} \, s^{-1}$. This
corresponds to an unabsorbed flux of $1.2 \times 10^{-11} \rm
\, erg \, cm^{-2} \, s^{-1}$ in the 0.2--10~keV band, and is
in good agreement with the HRC result above.   Both the
pulsed flux and the photon index of the pulsed emission are
in good agreement with previous results from Rosat (Finley et
al.\ 1993) and BeppoSAX (Mineo et al.\ (1999).

\begin{deluxetable}{lcc}
\tablecolumns{3}
\tablecaption{Properties of the Crab and 0540--69. \label{crab0540}}
\tablewidth{0pt}
\tablehead{                 & \colhead{Crab} & \colhead{0540--69}}
\startdata
Pulsar \.{E} (erg s$^{-1}$) & $4.7\times 10^{38}$  & $1.5\times 10^{38}$ \\
Total $L_X$                 & $3.4\times 10^{37}$  & $2.1\times 10^{37}$ \\
Compact nebula $L_{X}$      & $3.3\times 10^{37}$  & $1.7\times 10^{37}$ \\
Pulsed $L_{X}$              & $1.3\times 10^{36}$  & $0.4\times 10^{37}$ \\
Nebula size (pc)            & $1.5\times 1.5$      & $0.6\times 0.9$ \\
Pulsar offset (pc)          & 0.25                 & 0.09 \\
Braking index               & 2.51                 & 2.08 \\
\enddata
\tablecomments{X-ray luminosities, $L_{X}$, are quoted in erg~s$^{-1}$
for the 0.2--10~keV band.}
\end{deluxetable}

\subsubsection{Nebula spectrum} 

Fig.~\ref{ccimage} shows the pulsar flanked by the emission
from the nebula.  We accumulated spectra on both sides of the
pulsar, from $0.7\arcsec$ to $4.7\arcsec$ (left) and
$-0.7\arcsec$ to $-4.3\arcsec$ (right) and fitted both
spectra with a power law.  The flux from the nebula is
similar to that from the pulsar and thus the spectral
parameters are similarly constrained.  Here the fitted column
density is somewhat lower, which again is a systematic effect
of the low energy response below 1~keV, which now covers a
somewhat larger region on the device.  The spectral index of
the nebula, when fixing $N_{H}$ to the value used for the
pulsar, is slightly steeper than that of the pulsar, see
Table~\ref{specfits}.  The total absorbed flux in the two
flanks amounts to $1.47 \times 10^{-11} \rm \, erg \, cm^{-2}
\, s^{-1}$.  The pulsar flux at low phase may also be
associated with the nebula, which for the whole cycle amounts
to $1.04 \times 10^{-11} \rm \, erg \, cm^{-2} \, s^{-1}$.
Thus, the total absorbed flux for the compact nebula amounts
to $2.51 \pm 0.05 \times 10^{-11} \rm \, erg \, cm^{-2} \,
s^{-1}$.  The corresponding total nebular unabsorbed flux is
$4.8 \pm 0.05 \times 10^{-11} \rm \, erg \, cm^{-2} \,
s^{-1}$ in the 0.2--10~keV band.  This is good agreement with
the unabsorbed flux inferred from the HRC measurement.

\subsubsection{Outer remnant spectrum} 

Figure ~\ref{ccimage} shows that the pulsar and it's compact
nebula sits on top of some very extended excess emission, the
outer supernova remnant (Seward \& Harnden 1994).  We
accumulated a spectrum of this excess emission, in the region
between $-4.8\arcsec$ and $-30\arcsec$, and it turned out to
be extremely soft, i.e. the bulk of the emission is found to
be below 2~keV at a flux of approximately $9.3 \times
10^{-13} \rm \, erg \, cm^{-2} \, s^{-1}$.  The current
status of the ACIS response of front illuminated devices does
not allow us to do spectral fits in this area yet, but there
are line features evident in the  range between 0.7 and 1.8
keV and an overall resemblance to the spectrum observed in
E0102-72 (Hayashi et al.\ 1994; Gaetz et al.\ 2000).

\section{Conclusion}

Study of PSR B0540--69 with its compact nebula, and
comparison of its properties with those of the Crab and other
young pulsars, should provide insights into the physical
mechanisms for energy release in spin-powered pulsars
including the dynamics of the pulsar outflow.  In
Table~\ref{crab0540}, we compare the properties of PSR
B0540--69 with those of the Crab, assuming a distance to PSR
B0540--69 of 55~kpc.  

The fraction of the total spin-down power in x-ray emission
from the compact nebula and pulsar is 14\% for PSR B0540--69
and half that for the Crab.  The pulsed fraction of the total
(pulsar plus compact nebula) flux is 20\% for PSR B0540--69
compared to only 4\% for the Crab.  In addition, the pulse
shapes are quite different with the Crab showing a narrow
pulse while PSR B0540--69 has a broad, almost sinusoidal,
pulse.

PSR B0540--69 has lower spin-down power and the surrounding
nebula is less luminous.  As a general rule (Seward \& Wang
1985), the nebular luminosity varies with the pulsar
spin-down power, $\mbox{\.{E}}$, approximately as
$\mbox{\.{E}}^{1.4}$; the Chandra data are in rough
agreement.  The nebular volume for PSR B0540--69 is also
considerably less than that of the Crab, although it is
difficult to precisely define a ``size'' (the average FWHM is
0.5 pc and the average FW at background is 0.7 pc.  The FW at
background is given in Table~\ref{crab0540}).   Since the
nebula surrounding the pulsar is most likely due to
synchrotron emission and energized by the pulsar, the smaller
nebula likely reflects the lower spin-down power.  However,
the properties of the nebula may also depend on the
surrounding environment and the past history of the pulsar. 

The properties of the surrounding environment and also of the
progenitor may be important in determining the nature of the
outer remnant.  In PSR B0540--69, we detect thermal x-ray
emission from a surrounding remnant. This is in strong
contrast with the Crab pulsar, where no such remnant exists. 
The spectrum of the outer remnant of PSR B0540--69 appears to
be line driven and thus more of the category we see in young
oxygen-rich supernova remnants of the Magellanic Clouds
(Blair et al. 1999, Gaetz et al. 2000).

PSR B0540--69 and its surrounding nebula are a remarkable
parallel with the Crab Pulsar and its environment.  It is
almost certain that the same physical processes operate in
both systems.  However, further comparison shows several
distinct differences between the two systems, as revealed by
the observations described above.  Additional Chandra
observations would allow us to study PSR B0540--69 in detail;
making it possible to search for spectral structure in the
synchrotron nebula, as has recently been detected from the
Crab (Weisskopf et al.\ 2000).  Confrontation of detailed
models of the pulsar x-ray emission mechanism, pulsar wind,
synchrotron nebula, and the outer supernova remnant with
these observations should help advance our understanding of
spin-powered pulsars, their outflows, and the nature of their
progenitors.

\acknowledgments  

We thank Rob Cameron for providing information on the orbit
determination.  We gratefully acknowledge the efforts of the
Chandra team and support from NASA Chandra contract
NAS8-39073 and SAO grant SV1-61010.  PK acknowledges partial
support from NASA grant NAG5-7405.


\end{document}